\documentclass[A4paper,notitlepage,longbibliography,aps,twocolumn]{revtex4-1}
\usepackage{graphicx}
\graphicspath{{figures/}}
\usepackage{amsfonts}
\usepackage{amssymb}
\usepackage{mathrsfs}
\usepackage{soul}
\usepackage[usenames, dvipsnames]{color}
\usepackage[colorlinks=true,citecolor=blue,urlcolor  = purple]{hyperref}

\usepackage{tikz}

\usepackage[mathcal]{euscript}

\newcommand{\mcl}[1]{\mathcal{#1}}

\usepackage{amsmath}

\renewcommand{\d}{\mathrm{d}}
\newcommand{\Tr}{\operatorname{Tr}}

\newcommand{\ket}[1]{\left| #1 \right\rangle}

\newcommand{\ketbra}[2]{| #1 \rangle \langle #2 |}

\newcommand{\abs}[1]{\left| #1\right|}

\newcommand{\norm}[1]{\left\| #1 \right\|}

\newcommand{\kommentar}[1]{}

\usepackage{amsthm}

\newtheorem*{lemma*}{Lemma}
\newtheorem*{corollary*}{Corollary}
\theoremstyle{remark}

\usepackage{graphicx}

\usepackage{float}
\usepackage{color}
\definecolor{npurple}{rgb}{0.3,0,0.6}

\newcommand{\II}{\openone}
\newcommand{\I}{\openone}

\renewcommand{\H}{\mcl{H}}

\newcommand{\n}{\pmb{n}}
\renewcommand{\u}{\pmb{u}}
\renewcommand{\v}{\pmb{v}}
\newcommand{\e}{\pmb{e}}
\newcommand{\x}{\pmb{x}}

\renewcommand{\t}{\pmb{t}}

\renewcommand{\r}{\pmb{r}}

\definecolor{mygray}{gray}{0.6}

\begin{document}

\setstcolor{red}

\title{Quantum steering of Bell-diagonal states with generalized measurements}
\date{\today}

\author{H. Chau Nguyen}
\email{chau.nguyen@uni-siegen.de}
\affiliation{Naturwissenschaftlich-Technische Fakult\"at, Universit\"at Siegen,
Walter-Flex-Stra{\ss}e 3, 57068 Siegen, Germany}

\author{Otfried G\"{u}hne}
\email{otfried.guehne@uni-siegen.de}
\affiliation{Naturwissenschaftlich-Technische Fakult\"at, Universit\"at Siegen,
Walter-Flex-Stra{\ss}e 3, 57068 Siegen, Germany}

\begin{abstract}
The phenomenon of quantum steering in bipartite quantum systems can be 
reduced to the question whether or not the first party can perform 
measurements such that the conditional states on the second party can 
be explained by a local hidden state model. Clearly, the answer to this
depends on the measurements which the first party is able to perform. We 
introduce a local hidden state model explaining the conditional states
for all generalized measurements on Bell-diagonal states of two qubits. 
More precisely, it is known for the restricted case of projective 
measurements and Bell-diagonal states characterised by the correlation 
matrix $T$ that a local hidden state model exists if and only 
if $R_T= 2 \pi N_T \abs{\det (T)} \ge 1$, where $N_T$ is defined 
by an integral over the Bloch sphere. 
%via $N_T^{-1}=\int \d S (\n) [\n^T T^{-2} \n]^{-2}$.
For generalized measurements described by positive operator valued 
measures we construct a model if $R_T \ge 6/5$. Our work paves the 
way for a systematic study of steerability of quantum states with 
generalized measurements beyond the highly-symmetric Werner states. 
\end{abstract}

\maketitle

%%%%%%%%%%%%%%%%%%%%%%%%%%%%%%%%%%%%%%%%%%%%%%%%%%%%%%%%%%%%%%%%%%%%%
%Introduction
\section{Introduction}
Since its first description in 1935, the Einstein-Podolsky-Rosen (EPR) 
argument~\cite{Einstein1935a, Einstein48} has been at the center of 
many debates on the foundations of quantum mechanics and, at the same 
time, also a welling source of inspiration. Pondering on it, Bell 
introduced the concept of nonlocality~\cite{Bell1964a}; reconsidering 
the issue later, Werner and Primas defined modern notions of classical
correlations and entanglement~\cite{Primas83, Werner84, Werner1989a}. Both the theory of Bell 
nonlocality and of entanglement have played a crucial role in the 
development of modern quantum information theory. Based on ideas of 
Schr\"odinger, Wiseman and collaborators showed more recently that 
behind the EPR argument is yet another concept of quantum nonlocality, 
called quantum steering~\cite{Wiseman2007a, Schrodinger1935a}. This 
discovery since then has caused a new surge of research in quantum 
information theory. Characterization of quantum steering has been 
studied, many connections to different areas of quantum information 
were found and applications were established, for a recent review see 
Ref.~\cite{Uola2019a}.

%====================================================================================
%\textit{Quantum steerability.}

For explaining the notion of quantum steering, suppose Alice and Bob share 
a bipartite quantum state $\rho$ over the finite-dimensional Hilbert space 
$\H_A \otimes \H_B$ and Alice performs measurements on her side. The most 
general measurement with $n$ outcomes is described by a positive operator 
valued measure ($n$-POVM). Such a measure $E$ is a collection of $n$ positive 
operators, $E= \{E_i\}_{i=1}^{n}$, $E_i \ge 0$, fulfilling the normalization
$\sum_{i=1}^{n} E_i = \II_A$, where $\II_A$ is the identity operator on 
$\H_A$ (similarly, $\II_B$ will denote the identity operator acting on 
$\H_B$). 

After Alice performed the measurement and obtained the result, Bob's 
resulting states form the \emph{steering ensemble} of conditional
states, $\{\Tr_A [\rho (E_i \otimes \II_B)]\}_{i=1}^{n}$. If, however, 
the entanglement between the two parties is not sufficiently strong, 
the conditional states can be locally simulated by a so-called local 
hidden state model~\cite{Wiseman2007a}. The strategy of simulating 
goes as follows. Alice, instead of providing Bob with a part of the
entangled state, simply sends him an ensemble of states $\{\sigma_\lambda\}$ 
indexed by some variable $\lambda$ distributed according to a distribution $p(\lambda)$, 
known as \emph{local hidden states} 
(LHS).  

To simulate the steering corresponding to measurement $\{E_i\}_{i=1}^{n}$, 
Alice assigns state $\sigma_\lambda$ to outcome $i$ according to certain 
designed probability $G_i(\lambda)$. The functions $G=\{G_i\}_{i=1}^{n}$, thus 
satisfying $G_i(\lambda) \ge 0$ and $\sum_{i=1}^{n} G_i(\lambda)= 1$ 
for each $\lambda$, are referred to as \emph{response functions}. In 
this way Alice can simulate 
the steering ensemble by averaging over states associated to each 
outcome,
\begin{equation}
\Tr_{A}[\rho (E_i \otimes \II_B)]=  \sum_{\lambda} G_i(\lambda) p(\lambda) \sigma_{\lambda}.
\label{eq:unsteerable_def}
\end{equation}
Here the summation can be reformulated as an integration over an appropriate 
measure if required. Eq.~\eqref{eq:unsteerable_def} ensures that Bob, when 
performing state tomography conditioned on Alice's outcomes, obtains the 
same result as if Alice were steering his system by means of measurement 
$E$~\cite{Wiseman2007a}. If the design of the LHS ensemble $\{ \sigma_\lambda \}$ 
is such that for {\it all} measurements $E$  Alice can find a response function 
$G$ such that Eq.~\eqref{eq:unsteerable_def} holds, we say that the state is 
\emph{unsteerable}; otherwise it is \emph{steerable}.

This definition also extends naturally to the case where Alice is limited 
to certain subsets of measurements. Most often she is limited to performing 
projective measurements (or projector values measures, PVMs) only. In 
this case, for certain important entangled 
states LHS models exist, proving the states to be unsteerable~\cite{Wiseman2007a}. 
Some of these models can also be shown to be optimal, leading to an exact 
characterization of steering in these cases. These include the Werner 
states~\cite{Werner1989a,Wiseman2007a}, the isotropic states~\cite{Wiseman2007a} 
or general two-qubit states~\cite{Nguyen2018b}, where the solution for Bell-diagonal 
states is particularly simple \cite{Jevtic2015a,Nguyen2016a,Nguyen2016b}. If the most 
general measurements in quantum mechanics are taken into account, namely those 
described by general POVMs, the construction of LHS models is difficult. The  
most important known model is arguably the Barrett model, which was constructed 
for Werner states and isotropic states~\cite{Barrett2002a}. Contrary to the models 
for projective measurements, Barrett's model has never been demonstrated to be 
optimal; in fact evidence indicates that it is not~\cite{Nguyen2018a,Nguyen2018b}. 
Barrett's model also appeared to be ad hoc, seemingly limited to only Werner states 
and isotropic states, which are highly symmetric.

In this work, we generalize the Barrett construction of an LHS model to a more 
general family of two-qubit states, those which can be considered as a mixture 
of Bell states, or \emph{Bell-diagonal states}. Quantum steering of Bell-diagonal 
states when measurements are limited to projective measurements has been recently 
understood~\cite{Jevtic2015a, Nguyen2016a, Nguyen2016b}. With our new LHS model 
for POVMs  we can prove the unsteerability of a significant set of Bell-diagonal 
states. Our model is one of the first LHS models for POVMs on two-qubit states 
beyond the highly symmetric cases. In fact there is, to our knowledge, only one
other LHS model for POVMs, which is based on LHS models for projective 
measurements \cite{Hirsch2013a}. The model is, however, more limited and paradoxically 
appears to be more effective only for high-dimensional system rather than for 
two-qubit states. In our work we also explicitly demonstrate that our construction is
still non-optimal, as it fails to detect the unsteerability of some separable 
Bell-diagonal states. 

In the following, we start with introducing the Bloch four-vector representation 
of operators acting on qubit systems. After recalling the known results on quantum 
steering of Bell-diagonal states, we will describe the construction of our LHS model 
for POVMs. 

%%%%%%%%%%%%%%%%%%%%%%%%%%%%%%%%%%%%%%%%%%%%%%%%%%%%%%%%%%%%%%%%%%%%%
%Notations
\section{Bloch representation and Bell-diagonal states} 
For a qubit, any operator $X$ can be conveniently presented by a 
four-vector, 
%$\begin{pmatrix} 1 \\ \x \end{pmatrix}$
${x_0}\choose{\x}$, referred to as the Bloch representation. This 
is defined according to the expansion
\begin{equation}
X = \frac{1}{2}\sum_{i=0}^{3} x_i \sigma_i,
\end{equation}
where $\sigma_k$ for $k=0,1,2,3$ are the Pauli matrices extended 
to also include the identity matrix. In particular, projections 
or pure states are represented by 
%$\begin{pmatrix}1 \\ \n \end{pmatrix}$
${1}\choose{\n}$, with $\n$ being an unit vector. The set of 
pure states is thus a sphere, known as the Bloch sphere. 
Similarly, a two-qubit state $\rho$ can be most conveniently 
characterized by its Bloch tensor,
\begin{equation}
\rho = \frac{1}{4} \sum_{i,j=0}^{3} \Lambda_{ij} \sigma_i \otimes \sigma_j.
\end{equation}

The states that are diagonal in the Bell basis are those for which Alice's 
and Bob's reduced states are maximally mixed~\cite{Horodecki1996a}. The 
Bloch tensor $\Lambda$ can thus be written as
\begin{equation}
\Lambda = \begin{pmatrix}1  & 0  \\ 0 & T \end{pmatrix},
\end{equation}
where $T$ is s $(3 \times 3)$ matrix representing the correlation between 
them. Without loss of generality, we can assume that $T$ is diagonal. The 
reason is that the steering properties of a quantum state are invariant
under local unitary transformations acting on the states. These transformations
act as orthogonal transformations on the matrix $T$, and due to the singular
value decomposition this can be used to diagonalize $T$~\cite{Horodecki1996a}.

\section{Steering with projective measurements} 
To explain our construction, we first recall the optimal LHS 
model for Bell-diagonal states if measurements are limited to 
projective ones~\cite{Jevtic2015a,Nguyen2016a,Nguyen2016b}. A 
projective measurement on Alice's side corresponds to a pair 
of effects, which can be represented by Bloch vectors 
%$\begin{pmatrix}1 \\ \e_1 \end{pmatrix}$ 
${1}\choose{\e_1}$
and $ 1 \choose \e_2 $ with two unit vectors $\e_2=-\e_1$. 
As Alice gets outcomes $i \in \{1,2\}$, a direct calculation shows
that Bob's conditional states are 
\begin{equation}
\frac{1}{2} \begin{pmatrix}1 \\ T \e_i \end{pmatrix},
\label{eq:pvm_conditional}
\end{equation}
where we have used the fact that $T$ is symmetric, $T^T=T$.

Recall that the optimal ensemble of hidden states for a Bell-diagonal
state is defined by the following distribution on Bob's Bloch 
sphere~\cite{Jevtic2015a,Nguyen2016a}
\begin{equation}
p_J(\n)= \frac{N_T}{[\n^T T^{-2} \n]^2},
\label{eq:jevtic_ensemble}
\end{equation}
where $N_T$ is the normalization factor, given by
\begin{equation}
N_T^{-1} = \int \d S (\n) \frac{1}{[\n^T T^{-2} \n]^2},
\end{equation}
where the integral is taken over the surface of the Bloch 
sphere with $\d S$ being the surface measure. The response 
function for \emph{projective measurements} is chosen 
as~\cite{Jevtic2015a}
\begin{equation}
G^J_i (\n)= \Theta (\e^T_i T^{-1} \n),
\label{eq:jevtic_response}
\end{equation}
where $\Theta$ is the Heaviside step function.
 
The conditional states produced by this LHS model can be 
computed to be
\begin{equation}
\int  \d S (\n) \frac{N_T}{[\n^T T^{-2} \n]^2} \Theta (\e^T_i T^{-1} \n) \begin{pmatrix}1 \\ \n  \end{pmatrix}=
 \frac{1}{2} \begin{pmatrix} 1 \\ R_T T \e_i \end{pmatrix},
\label{eq:pvm_simulated}
\end{equation}
where 
\begin{equation}
R_T= 2 \pi   N_T \abs{\det(T)}.
\label{eq:tstates_critical}
\end{equation}
Comparing this with the conditional states due to the actual measurements 
at Alice's side in Eq.~\eqref{eq:pvm_conditional}, one sees that 
$R_T \ge 1$ suffices to deduce that the corresponding Bell-diagonal 
state is unsteerable~\cite{Jevtic2014a,Jevtic2015a}: If $R_T = 1$
the LHS model simulates exactly the state, while for $R_T > 1$ the 
LHS model allows even to simulate a state with stronger correlations.

The described construction of the LHS ensemble was later proven to be optimal~\cite{Nguyen2016a,Nguyen2016b}, thus this is also a necessary 
condition for the Bell-diagonal state to be unsteerable. In fact, the 
quantity $R_T$ has been generalized for general states, where it is 
referred to as the critical radius of the state~\cite{Nguyen2018a}. 

%Positive operator valued measures

\section{Steering with positive operator valued measures} 
We now consider the case where Alice makes a POVM. It is known 
that in studying steerability of qubits, it is sufficient to 
consider POVMs consisting of four rank-$1$ effects, as these 
are the extremal POVMs~\cite{Barrett2002a,Dariano2005a}. A 
range-$1$ POVM of $4$ outcomes is given by effects 
$\alpha_i {{1}\choose{\e_i}}$ with $0 \le \alpha_i \le 1$ 
such that
\begin{equation}
\sum_{i=1}^{4} \alpha_i \begin{pmatrix}1 \\ \e_i \end{pmatrix} = \begin{pmatrix}
2 \\ \pmb{0}
\end{pmatrix},
\label{eq:normalisation_povm}
\end{equation}
where one should recall that 
$2 \choose \pmb{0}$ represents the identity operator. Corresponding 
to the measurement outcome $i$, the conditional state on Bob's side 
is
\begin{equation}
\frac{\alpha_i}{2}  \begin{pmatrix}1 \\ T \e_i \end{pmatrix}.
\label{eq:povm_conditional}
\end{equation} 
Our purpose is to construct a possible LHS model to model these 
conditional ensembles of states. The probability distribution of the
LHS ensemble is still defined by Eq.~\eqref{eq:jevtic_ensemble}. To 
define the response function, we first denote
\begin{equation}
g_i(\n)= \frac{1}{2} \left( 1 + \frac{\e_i^T T^{-1} \n}{\norm{T^{-1} \n}} \right) \Theta (\e^T_i T^{-1} \n),
\label{eq:pre_response_function}
\end{equation}
and 
\begin{equation}
 c_i = \int \d S (\n) p_J(\n) g_i (\n).
\end{equation}
Note that
\begin{equation}
0 \le c_i \le \int \d S (\n) p_J(\n) \Theta (\e^T_i T^{-1} \n) = \frac{1}{2},
\end{equation} 
where the last identity is because $p_J(\n)=p_J(-\n)$. 

Inspired by Barrett's construction of an LHS model for the Werner 
states~\cite{Barrett2002a}, we construct the response function 
for the Bell-diagonal states by
\begin{equation}
G_i(\n)= \alpha_i g_i (\n) + \gamma_i \left[ 1 - \sum_{j=1}^{n} \alpha_j g_j(\n)\right]
\label{eq:barrett_response}
\end{equation}
with 
\begin{equation}
\gamma_i=  \frac{\alpha_i \left( \frac{1}{2} - c_i\right) }{1-\sum_{k=1}^{n} \alpha_k c_k}.
\end{equation}
Recalling from Eq.~\eqref{eq:normalisation_povm} that $\sum_{i=1}^{n} \alpha_i=2$, 
one can deduce that $\sum_{i=1}^{n} \gamma_i=1$. As a consequence, one can also 
easily verify that $\sum_{i=1}^{n} G_i(\n)=1$. Since $c_i \le \frac{1}{2}$ we have 
that $\gamma_i \ge 0$. So, in order to show that $G_i(\n)$ is positive, it is 
sufficient to show that
\begin{equation}
\sum_{i=1}^{n} \alpha_i g_i (\n) \le 1. 
\label{eq:subnormalisation}
\end{equation}
This is seen when one notices
\begin{equation}
\sum_{i=1}^{n} \alpha_i g_i (\n) \le \frac{1}{2} \sum_{i=1}^{n} \alpha_i\left( 1 + \frac{\e_i^T T^{-1} \n}{\norm{T^{-1} \n}} \right)=1,
\end{equation}
where we have used Eq.~\eqref{eq:normalisation_povm} to identify 
$\sum_{i=1}^{n} \alpha_i = 2$ and $\sum_{i=1}^{n} \alpha_i \e_i = \pmb{0}$. 
To summarize, $G_i(\n)$ is indeed a valid response function.

Let us digress for a moment to comment on the rationale behind the 
construction in Eq.~\eqref{eq:pre_response_function} and 
Eq.~\eqref{eq:barrett_response}. We first notice that the 
conditional state in Eq.~\eqref{eq:povm_conditional} for POVMs 
is very similar to that for PVMs in Eq.~\eqref{eq:pvm_conditional}. 
The only difference is the prefactor $\alpha_i$ in 
Eq.~\eqref{eq:povm_conditional}. One thus might attempt to modify 
the response function for projective measurements in 
Eq.~\eqref{eq:jevtic_response} by a multiplicative factor 
$\alpha_i$. Unfortunately, since a POVM has four effects
this does not result in a valid response function, 
in contrast to projective measurements, since the summation 
of the response functions over the outcomes exceeds one. The 
strategy is to soften the response function in Eq.~\eqref{eq:jevtic_response} 
first to the form in Eq.~\eqref{eq:pre_response_function} by multiplication
with a linear function. The linear softener in Eq.~\eqref{eq:pre_response_function} 
was chosen to guarantee the subnormalization for all measurements in
Eq.~\eqref{eq:subnormalisation}. The subnormalization is then corrected 
by an additive factor in Eq.~\eqref{eq:barrett_response} such that 
normalization is guaranteed and we obtain a valid response function. 

Having shown that the response functions in Eq.~\eqref{eq:barrett_response} 
are valid, it remains to demonstrate that the simulated states match the 
steered states as required by the definition of unsteerability in
Eq.~\eqref{eq:unsteerable_def}. We will now show that the conditional 
state Alice can simulate is given by
\begin{equation}
\int \d S (\n) p_J(\n) G_i (\n) 
\begin{pmatrix} 1 \\ \n \end{pmatrix}
=\frac{\alpha_i}{2}  
\begin{pmatrix}1 \\ \frac{5R_T}{6} T \e_i \end{pmatrix},
\label{eq:tsates_povm_simulated}
\end{equation} 
with $R_T$ defined by Eq.~\eqref{eq:tstates_critical}. In a similar 
manner with the elaboration of Eq.~\eqref{eq:pvm_simulated}, we can 
deduce that a Bell-diagonal state is unsteerable with POVMs if $R_T \ge 6/5$. 

The derivation of Eq.~\eqref{eq:tsates_povm_simulated} is as follows. By construction~\eqref{eq:barrett_response} we already have,
\begin{equation}
\int \d S (\n) p_J(\n) G_i(\n)= \frac{\alpha_i}{2}, 
\label{eq:scalar_part}
\end{equation}
independent of $T$. We now compute
\begin{equation}
\int \d S (\n) p_J(\n) G_i (\n) \n = \alpha_i \t_i - \gamma_i \sum_{k=1}^{n} \alpha_k \t_k,
\end{equation}
where we denote
\begin{equation}
\t_i = \int \d S (\n) p_J(\n) g_i (\n) \n.
\end{equation}

To compute the $\t_i$, we decompose them as
\begin{equation}
\t_i=\u_i + \v_i,
\label{eq:t_i_0}
\end{equation}
where
\begin{align}
\u_i & = \frac{N_T}{2} \int  \frac{\d S (\n)}{[\n^T T^{-2} \n]^2} \Theta (\e_i^T T^{-1} \n) \n
\label{ui-def}
\\
\v_i & =  \frac{N_T}{2} \int  \frac{\d S (\n)}{[\n^T T^{-2} \n]^{5/2}} (\e_i^T T^{-1} \n) \Theta (\e^T_i T^{-1} \n) \n.
\label{vi-def}
\end{align}

The computation of $\u_i$ has been performed in Ref.~\cite{Jevtic2015a} using 
spherical coordinates.  A simpler and coordinate-independent technique to 
compute $\u_i$ was introduced in Ref.~\cite{Nguyen2018a}. In order to be 
self-contained and for pedagogical reasons, we present here the computation 
of both $\u_i$ and $\v_i$ using the technique introduced in 
Ref.~\cite{Nguyen2018a}. 

\begin{itemize}
\item[(i)] To compute $\u_i$, we consider
\begin{equation}
\tilde{\u}_i = \int \d V (\r) e^{-\frac{1}{2} \r^T T^{-2} \r} \Theta (\e_i^T T^{-1} \r) \r,
\label{eq:u_i_0}
\end{equation}
where $V$ is the volume measure and the integral is taken 
over the whole three-dimensional space of $\r$. 

Let $\r = r \n$, with $\n$ being a unit vector. Denoting the 
surface measure of the unit sphere by $\d S$, we have
\begin{align}
\tilde{\u}_i &= \int \d S (\n) \Theta (\e_i^T T^{-1} \n) \n \int_{0}^{+\infty} \!\!\!\!\!\!\!\! \d r r^3 e^{-\frac{r^2}{2} \n^T T^{-2} \n} \nonumber \\
&=\int  \frac{\d S (\n)}{[\n^T T^{-2} \n]^2} \Theta (\e_i^T T^{-1} \n) \n \int_0^{+\infty} \!\!\!\!\!\!\!\! \d x x^3 e^{-\frac{x^2}{2}} \nonumber  \\
& = 2 \int  \frac{\d S (\n)}{[\n^T T^{-2} \n]^2} \Theta (\e_i^T T^{-1} \n) \n.
\label{eq:u_i_tilde}
\end{align}
Comparing this with Eq.~(\ref{ui-def}) we can conclude that $\u_i=N_T \tilde{\u}_i/4$.

On the other hand, by applying the transformation $\r=T\r'$ to the integral in
Eq.~\eqref{eq:u_i_0}, $\tilde{\u}_i$ can be computed explicitly, 
\begin{align}
\tilde{\u}_i &= \abs{\det(T)} \int \d V (\r) e^{-\frac{1}{2} \r^T \r} \Theta (\e_i^T \r) T \r \nonumber \\
&= \abs{\det(T)}  T \e_i \int_{-\infty}^{+\infty} \!\!\!\!\!\!\!\! \d x \d y \d z e^{-\frac{1}{2} (x^2 + y^2 + z^2)} \Theta (z) z \nonumber \\
&= \abs{\det(T)}  2 \pi T \e_i.
\end{align}
Therefore we have:
\begin{equation}
\u_i = \abs{\det(T)} \frac{ \pi N_T}{2} T \e_i.
\label{eq:u_i}
\end{equation}

\item[(ii)] The evaluation of $\v_i$ is similar. We consider
\begin{equation}
\tilde{\v}_i = \int \d V (\r) e^{-\frac{1}{2} \r^T T^{-2} \r} (\e_i^T T^{-1} \r) \Theta (\e_i^T T^{-1} \r) \r.
\label{eq:v_i_0}
\end{equation}
With the same notation as in Eq.~\eqref{eq:u_i_tilde}, we have
\begin{align}
\tilde{\v}_i &= \int \d S (\n) (\e_i^T T^{-1} \n) \Theta (\e_i^T T^{-1} \n) \n \int_{0}^{+\infty} \!\!\!\!\!\!\!\! \d r r^4 e^{-\frac{r^2}{2} \n^T T^{-2} \n} \nonumber \\
&=\int  \frac{\d S (\n)}{[\n^T T^{-2} \n]^{5/2}} (\e_i^T T^{-1} \n) \Theta (\e_i^T T^{-1} \n) \n \int_0^{+\infty} \!\!\!\!\!\!\!\! \d x x^4 e^{-\frac{x^2}{2}} \nonumber \\
& = \frac{3 \sqrt{\pi}}{\sqrt{2}} \int  \frac{\d S (\n)}{[\n^T T^{-2} \n]^{5/2}} (\e_i^T T^{-1} \n) \Theta (\e_i^T T^{-1} \n) \n.
\end{align}
So we obtain $\v_i=N_T \tilde{\v}_i/(3 \sqrt{2 \pi})$.

On the other hand, by applying the transformation $\r=T\r'$ to the integral Eq.~\eqref{eq:v_i_0}, $\tilde{\v}_i$ can also be computed explicitly, 
\begin{align}
\tilde{\v}_i &= \abs{\det(T)} \int \d V (\r) e^{-\frac{1}{2} \r^T \r} (\e_i^T \r) \Theta (\e_i^T \r)  T \r \nonumber \\
&= \abs{\det(T)}  T \e_i \int_{-\infty}^{+\infty} \!\!\!\!\!\!\!\! \d x \d y \d z e^{-\frac{1}{2} (x^2 + y^2 + z^2)} \Theta (z) z^2 \nonumber \\
&= \abs{\det(T)}  \pi \sqrt{2 \pi} T \e_i.
\end{align}
Therefore
\begin{equation}
\v_i = \abs{\det(T)} \frac{\pi N_T}{3}   T \e_i.
\label{eq:v_i}
\end{equation}
\end{itemize}

Inserting Eq.~\eqref{eq:u_i} and Eq.~\eqref{eq:v_i} into 
Eq.~\eqref{eq:t_i_0}, we obtain 
\begin{equation}
\t_i = \abs{\det(T)} \frac{5 \pi N_T}{6}   T \e_i.
\end{equation}
In particular, since $\sum_{i=1}^{4} \alpha_i \e_i = 0$ due 
to Eq.~\eqref{eq:normalisation_povm}, we have
\begin{equation}
\sum_{i=1}^{n} \alpha_i \t_i= 0.
\end{equation}
Thus
\begin{equation}
\int \d S (\n) p_J(\n) G_i (\n) \n = \abs{\det(T)} \frac{5 \pi N_T}{6}   T \e_i.
\label{eq:vector_part}
\end{equation}
Together, Eq.~\eqref{eq:vector_part} and Eq.~\eqref{eq:scalar_part} deliver 
the announced result Eq.~\eqref{eq:tsates_povm_simulated}.

\section{Examples}

Let us now consider several special cases of the condition 
$R_T \ge 6/5$ for Bell-diagonal states to be unsteerable 
with POVMs. 

\subsection{Werner states} 
For two qubits, the Werner states are given by
\begin{equation}
W_p= p\ketbra{\psi_-}{\psi_-} + (1-p) \frac{\I}{2} \otimes \frac{\I}{2},
\end{equation}
where $\ket{\psi_-}=(\ket{01}-\ket{10})/\sqrt{2}$ and $0 \le p \le 1$. 
The Werner states are certainly a special Bell-diagonal state, with 
the critical radius $R_{W_p}=1/(2p)$~\cite{Nguyen2018a}.The condition $R_{W_p} \ge 1$ for 
it to be unsteerable with PVMs reduces to $p \le 1/2$, and the 
condition $R_{W_p} \ge 1$ for it to be unsteerable with POVMs 
reduces to $p \le 5/12$, both of which are 
well-known~\cite{Werner1989a,Barrett2002a,Wiseman2007a}. 

%%%%%%%%%%%%%%%%%%%%%%%%%%%%%%%%%%%%%%%%%%%%%%%%%%%%%%%%%%%%5
\begin{figure}[t!]
\begin{center}
\includegraphics[width=0.7\columnwidth]{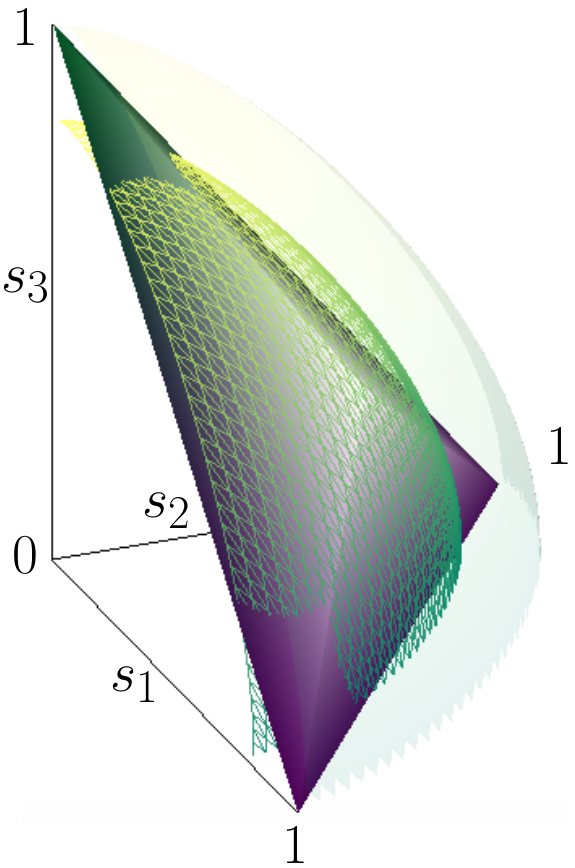}
\end{center}
\caption{Separability, steerability with PVMs and POVMs of 
Bell-diagonal states. A general Bell-diagonal state is, up 
to local unitary transformations described by three parameters 
$(s_1, s_2, s_3)$, which, without loosing generality, can be 
assumed to be positive. The states with all $s_i$ equal correspond 
to the Werner states. Using the original Barrett construction and 
the fact that separable states have an LHS model, one can see that 
the states in the volume characterized by the green-violet surface 
have an LHS model. The surface of states for which our model works ($R_T=6/5$)
is shown as a yellow-green grid. The border of unsteerable states 
with projective measurements ($R_T=1$) is shown as a diaphanous surface. See the 
text for discussion.}
\label{fig:tstates}
\end{figure}
%%%%%%%%%%%%%%%%%%%%%%%%%%%%%%%%%%%%%%%%%%%%%%%%%%%%%%%%%%%%%%%%%%%%

\subsection{Axially symmetric Bell-diagonal states} 
For the case of Bell-diagonal states with axial symmetry, $T=\operatorname{diag}(-s,-s,-t)$, 
the critical radius $R_T$ can be evaluated explicitly as ~\cite{Nguyen2018a}
\begin{equation}
R_T= \frac{1}{\abs{t}} \frac{1}{1+x^2 \frac{\operatorname{arctg} \sqrt{x^2-1} }{\sqrt{x^2-1}}},
\end{equation}
where $x=s/t$ and $\sqrt{x^2-1}$ can take purely imaginary values when $\abs{x} \le 1$. With this, we find the axially symmetric Bell diagonal state is unsteerable with PVMs if and only if, $\abs{t}^{-1} \ge 1+x^2/ \sqrt{x^2-1} \operatorname{arctg} (\sqrt{x^2-1})$, and unsteerable with POVMs if $5/6 \abs{t}^{-1} \ge 1+x^2/ \sqrt{x^2-1} \operatorname{arctg} (\sqrt{x^2-1})$.

\subsection{General Bell-diagonal states} 
Let us finally discuss general Bell-diagonal states with 
$T=-\operatorname{diag}(s_1,s_2,s_3)$. In the space of 
$(s_1,s_2,s_3)$, the separable Bell diagonal states 
constitute an octahedron~\cite{Horodecki1996a}, which 
is symmetric under the three reflections $s_i \to -s_i$. 
Moreover $R_T$ depends only on the absolute values of 
$s_i$. We thus can concentrate on the steerability 
and separability of the Bell-diagonal states in the 
positive octant $s_1 \ge 0$, $s_2 \ge 0$, $s_2 \ge 0$.

This situation is depicted in Fig.~\ref{fig:tstates}. 
In the positive octant, the border separating separable 
states from entangled state is a triangle with vertices 
$(1,0,0)$, $(0,1,0)$, $(0,0,1)$. Note that separable states 
are certainly unsteerable with POVMs. The original Barrett 
model for Werner states shows that the state corresponding 
to $(5/12,5/12,5/12)$ is also unsteerable with POVMs. 
Taking the convex hull of this point $(5/12,5/12,5/12)$ 
with those of the separable states, we obtain a polytope of
states which are definitely unsteerable with POVMs. The 
faces of this polytopes are presented in Fig.~\ref{fig:tstates} 
together with the surface $R_T=6/5$. The figure illustrates 
that our model demonstrates a significant new volume of 
Bell-diagonal states to be unsteerable with POVMs, in 
comparison to what has been known in the literature. 

It should be noted, however, that our model does not work for all
the states in the polytope mentioned above. In fact, the introduced 
model does not demonstrate the unsteerability of certain separable 
states, indicating that it is not optimal. Extrapolating to the 
 Barrett's original construction, one can expect that this 
well-known model is also not optimal. This is compatible with 
the conjecture that POVMs and PVMs are equivalent in quantum 
steering with the two-qubit Werner states, which is supported 
by some numerical evidence \cite{Werner2014a,Nguyen2018a,Nguyen2018b}.

\section{Conclusion} 
At first sight, Barrett's construction of an LHS model for 
Werner states~\cite{Barrett2002a} appears to be ad hoc. Our 
generalization illustrates a certain rational reasoning behind 
the construction. However, the very fact that 
Eq.~\eqref{eq:tsates_povm_simulated} holds only came out though a 
rather complicated computation. The question whether this identity 
is a mathematical coincidence or there is a deeper mathematical 
reason underpinned it is an interesting question. We expect 
that the answer to this question can possibly lead to LHS models 
for a much more general class of bipartite states, shedding light 
on the role of POVMs in quantum steering and quantum nonlocality. 

%%%%%%%%%%%%%%%%%%%%%%%%%%%%%%%%%%%%%%%%%%%%%%%%%%%%%%%%%%%%%%%%%%%%%%%%%%%%
\begin{acknowledgements}
This work was supported by the DFG and the ERC (Consolidator Grant
683107/TempoQ). 
%H.~C.~N. also acknowledges 
%the support by the Vietnam National Foundation for Science and Technology 
%Development (NAFOSTED) under grant number 103.02-2015.48.
\end{acknowledgements}

\bibliography{quantum-steering}
%%%%%%%%%%%%%%%%%%%%%%%%%%%%%%%%%%%%%%%%%%%%%%%%%%%%%%%%%%%%%%%%%%%%%

\end{document}